\documentstyle[preprint,tighten,aps,amstex,graphicx,times,floats]{revtex}

\newcommand{\centeron}[2]{{\setbox0=\hbox{#1}\setbox1=\hbox{#2}\ifdim
                             \wd1>\wd0\kern.5\wd1\kern-.5\wd0\fi \copy0
                             \kern-.5\wd0\kern-.5\wd1\copy1\ifdim\wd0>\wd1
                             \kern.5\wd0\kern-.5\wd1\fi}}

\newcommand{\gtap}{\>\centeron{\raise.35ex\hbox{$>$}}
                     {\lower.65ex\hbox{$\sim$}}\>}
\newcommand{\gsim}{\mathrel{\gtap}}

\newcommand{\ltap}{\>\centeron{\raise.35ex\hbox{$<$}}
                     {\lower.65ex\hbox{$\sim$}}\>}
\newcommand{\lsim}{\mathrel{\ltap}}

\begin{document}

\thispagestyle{empty}
 
\marginparwidth 1.cm
\setlength{\hoffset}{-1cm}
\newcommand{\mpar}[1]{{\marginpar{\hbadness10000%
                      \sloppy\hfuzz10pt\boldmath\bf\footnotesize#1}}%
                      \typeout{marginpar: #1}\ignorespaces}
\def\mda{\mpar{\hfil$\downarrow$\hfil}\ignorespaces}
\def\mua{\mpar{\hfil$\uparrow$\hfil}\ignorespaces}
\def\mla{\marginpar[\boldmath\hfil$\rightarrow$\hfil]%
                   {\boldmath\hfil$\leftarrow $\hfil}%
                    \typeout{marginpar: $\leftrightarrow$}\ignorespaces}

\def\ba{\begin{eqnarray}}
\def\ea{\end{eqnarray}}
\def\bq{\begin{equation}}
\def\eq{\end{equation}}

\renewcommand{\abstractname}{Abstract}
\renewcommand{\figurename}{Figure}
\renewcommand{\refname}{Bibliography}

\newcommand{\eg}{{\it e.g.}\;}
\newcommand{\ie}{{\it i.e.}\;}
\newcommand{\etal}{{\it et al.}\;}
\newcommand{\ibid}{{\it ibid.}\;}

\newcommand{\mx}{M_{\rm SUSY}}
\newcommand{\pt}{p_{\rm T}}
\newcommand{\et}{E_{\rm T}}
\newcommand{\del}{\varepsilon}
\newcommand{\sla}[1]{/\!\!\!#1}
\newcommand{\fb}{\;{\rm fb}}
\newcommand{\pb}{\;{\rm pb}}
\newcommand{\mev}{\;{\rm MeV}}
\newcommand{\gev}{\;{\rm GeV}}
\newcommand{\tev}{\;{\rm TeV}}
\newcommand{\abi}{\;{\rm ab}^{-1}}
\newcommand{\fbi}{\;{\rm fb}^{-1}}

\newcommand{\lsusy}{\lambda'_{\rm SUSY}}
\newcommand{\llq}{\lambda'_{\rm LQ}}
\newcommand{\msbar}{\overline{\rm MS}}
\newcommand{\met}{E\hspace{-0.45em}|\hspace{0.1em}}

\newcommand{\SP}{\scriptscriptstyle}
\newcommand{\stl}{\tilde{t}_{\SP L}}
\newcommand{\str}{\tilde{t}_{\SP R}}
\newcommand{\ste}{\tilde{t}_1}
\newcommand{\stz}{\tilde{t}_2}
\newcommand{\st}{\tilde{t}}
\newcommand{\gt}{\tilde{g}}
\newcommand{\sle}{\tilde{\tau}_1}
\newcommand{\slz}{\tilde{\tau}_2}
\newcommand{\che}{\tilde{\chi}^\pm_1}
\newcommand{\cpe}{\tilde{\chi}^+_1}
\newcommand{\cme}{\tilde{\chi}^-_1}
\newcommand{\nne}{\tilde{\chi}^0_1}
\newcommand{\nnz}{\tilde{\chi}^0_2}
\newcommand{\mse}{m_{\tilde{t}_{\SP 1}}}
\newcommand{\msz}{m_{\tilde{t}_{\SP 2}}}
\newcommand{\mst}{m_{\tilde{t}}}
\newcommand{\mg}{m_{\tilde{g}}}
\newcommand{\ms}{m_{\tilde{q}}}
\newcommand{\mt}{m_t}
\newcommand{\mheavy}{m_{\rm heavy}}
\newcommand{\mle}{m_{\tilde{\tau}_{\SP 1}}}
\newcommand{\mce}{m_{\tilde{\chi}^+_{\SP 1}}}
\newcommand{\mne}{m_{\tilde{\chi}^0_{\SP 1}}}

\preprint{
\font\fortssbx=cmssbx10 scaled \magstep2
\hbox to \hsize{
\hfill\vtop{\hbox{\today}} }
}

\title{High-Energy Neutrino Astronomy: Opportunities For Particle Physics}

\author{Dan Hooper}

\address{
Theoretical Astrophysics, University of Oxford, OX1 3RH Oxford, England, UK}

\maketitle

\begin{abstract}

In this article, based on the talk given at the Cracow Epiphany Conference on Astroparticle Physics, I discuss some of the opportunities provided by high-energy and ultra-high energy neutrino astronomy in probing particle physics beyond the standard model. Following a short summary of current and next generation experiments, I review the prospects for observations of high-energy neutrino interactions, searches for particle dark matter, and measurements of absolute neutrino masses, lifetimes and pseudo-Dirac mass splittings. 

\end{abstract}

\section{Introduction}

In the past few years, a new scale of high-energy neutrino experiment has been constructed. These experiments, including AMANDA-II and RICE, have opened a new window through which we can study the sky and the objects in it. Soon to follow are the next generation of high-energy neutrino experiments, including Antares in the Mediterranean, IceCube at the South Pole and a host of ultra-high energy projects such as the Pierre Auger observatory, ANITA and EUSO. 

These new experiments will be adept at probing a variety of astrophysical puzzles, including the origin of the highest energy cosmic rays, and possibly related, the inner workings of active galactic nuclei, gamma-ray bursts and other astrophysical accelerators. Neutrino astronomy is not limited to astrophysical endeavors, however. These experiments will observe neutrino interactions at center-of-mass energies far beyond those possible in any planned accelerator. They can also observe neutrinos which have travelled over hundreds, or thousands, of Megaparsecs, providing a new probe of neutrino stability, pseudo-Dirac mass splittings and even quantum gravity.

It is a fair criticism that, in contrast to accelerator-based physics, the luminosity of cosmic neutrinos is small, and not well known. For this reason, precision particle physics is now, and will remain for the foreseeable future, within the realm of colliders. In this talk, I will explore many of the particle physics scenarios which can be tested without collider-level precision. In particular, I will discuss models with strong interactions above the TeV scale, including models of low scale quantum gravity (ADD, Randall-Sundrum, string excitations, etc.). I will also talk about searches for particle dark matter, both at the TeV and Grand Unified scales. Thirdly, I will discuss probes of neutrino properties, such as absolute neutrino masses, neutrino decays, and pseudo-Dirac neutrino mass splittings.

\section{The Experiments}

AMANDA-II \cite{amanda}, located a mile under the South Pole, consists of 19 strings of photomultipliers (PMs) buried deep in the Antarctic ice. It is sensitive to neutrino induced muons above 30 to 50 GeV, has an effective area of 50,000 square meters, and has been taking data for over three years in its current configuration, although only a small fraction of this data has been unblinded thus far. The PMs are sensitive to optical Cerenkov light from both muon tracks and hadronic or electromagnetic showers, allowing for observation of all neutrino flavors. Unlike experiments sensitive only at much higher energies, AMANDA-II has been calibrated using the flux of atmospheric neutrinos with energies of 500 GeV to 100 TeV.

Antares \cite{antares} uses a similar approach in deep ocean water, about 40 kilometers off of the French coast.  When completed, Antares will consist of 12 strings, yielding an effective area of roughly 0.1 square kilometers. In addition to a larger effective area than AMANDA, Antares will also be sensitive at lower energies, quoting a muon energy threshold near 10 GeV. 

The IceCube experiment \cite{icecube} represents a major step forward in high-energy neutrino astronomy. With a full square kilometer of effective area, IceCube will reach the required sensitivity to probe many puzzles in both astrophysics and particle physics and will be the focus of much of this talk. Beginning installation next year, IceCube is scheduled to be completed in 2009. During this time period, any installed strings will accumulate data prior to the experiment's completion, allowing for a full square kilometer-year of exposure or more prior to the end of IceCube's construction.

In addition to a large volume, IceCube will have an angular resolution of less than $1^{\circ}$ for muon tracks, and $10^{\circ}$ for shower events. Energy resolution for muons is also quite good, better than 30\% in the log of energy. For showers, energy resolution scales linearly with energy and is better than 20\%. Energy resolution is critical, as above the PeV scale, the atmospheric neutrino background drops dramatically, making very high-energy events particularly interesting. 

\begin{figure}[t]
\begin{center}
\includegraphics[width=7.0cm,angle=0]{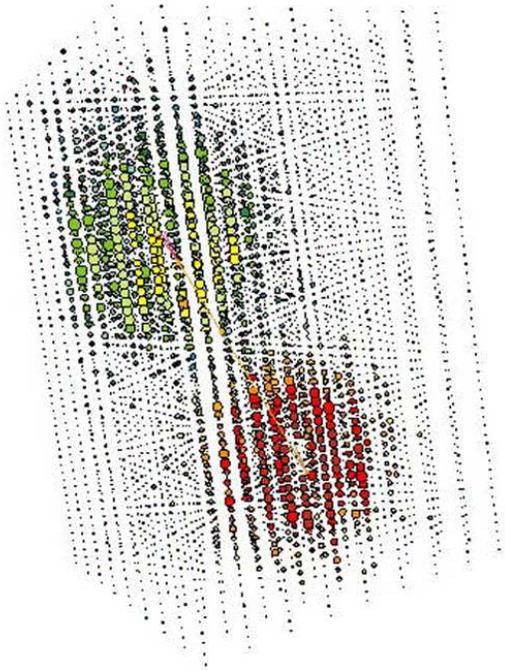}
\end{center}
\vspace*{0mm}
\caption[]{\label{doublebang}
A simulation of a ``double bang'' tau neutrino event in IceCube.}
\end{figure}

In addition to muon tracks and hadronic or electromagnetic showers, IceCube can also distinguish events unique to tau neutrinos \cite{doublebang}. The first of these event-types is the so-called, ``double bang'' signature.  In such an event, a multi-PeV tau neutrino enters the detector volume and interacts via charged current, producing a shower and a tau lepton. This tau lepton travels through the detector, but decays before leaving the effective volume, thus producing a second shower, or ``bang'' (see figure\ref{doublebang}). A second event topology unique to tau neutrinos is the ``lollipop'' event. In this case, the first bang of a ``double bang'' event occurs outside of the detector, and the resulting tau lepton travels into the effective volume, where it decays.  In this picture, the tau track is the stick of the lollipop and the decay shower the candy.  This type of event can be important at energies well above the few PeV range where double bangs are observed, due to the multi-kilometer tau decay distance for such events.

In addition to optical Cerenkov experiments, with 10-100 GeV energy thresholds, there are efforts to design and build detectors optimized for considerably higher energies. In particular, experiments such as RICE \cite{rice} at the South Pole and balloon-based ANITA \cite{anita}, observe EeV-scale neutrinos by Giga-Hertz radio emission produced from the excess electrons in neutrino induced showers \cite{radio}. Observations at very high energies with acoustic techniques have also been explored \cite{acoustic}.

Ultra-high energy cosmic ray experiments can also be excellent neutrino detectors at the highest energies. Experiments such as AGASA and Auger observe neutrinos as quasi-horizontal showers, which penetrate more deeply into the atmosphere than a hadronic (or photon) primary. In the future, satellite based cosmic ray experiments, such as EUSO, will be very sensitive EeV neutrino observatories.

There are many interesting possibilities for high-energy neutrino astronomy beyond IceCube, Auger, ANITA and EUSO. These include the proposed SALSA experiment, which would use large salt domes as a radio Cerenkov medium \cite{salsa}. Extensions of IceCube, dedicated to ultra-high energy neutrinos, have also been proposed \cite{hypercube}.

For a review of high-energy neutrino experimentation, as well as astrophysical sources, see Ref.~\cite{review}.

\section{High-Energy Neutrino Interactions}

Neutrino-nucleon interaction cross sections can be enhanced at high-energies in a variety of particle physics scenarios.  These include scenarios with large extra dimensions (ADD) \cite{add}, warped extra dimensions (Randall-Sundrum) \cite{rs}, or string excitations \cite{veneziano}, as well as microscopic black hole production \cite{bh} and even standard model electroweak instanton induced processes \cite{instanton}. I will not discuss the details of any of these models in this talk, focusing rather on the phenomenological aspects relevant to neutrino astronomy.

\begin{figure}[t]
\begin{center}
\includegraphics[width=10.0cm,angle=0]{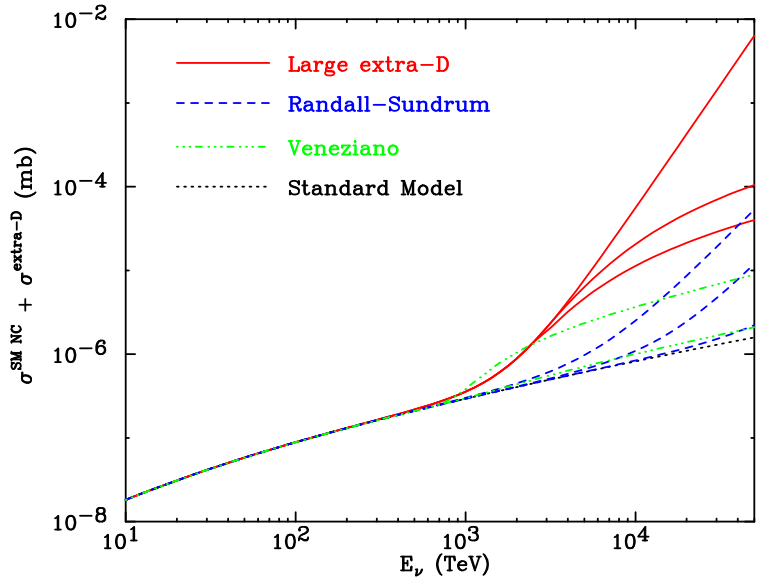}
\end{center}
\vspace*{0mm}
\caption[]{\label{sigmagravity}
Neutrino-nucleon cross sections for a variety of TeV-scale quantum gravity models \cite{gravity}.}
\end{figure}

Using cosmic neutrinos to study high-energy particle interactions can be a viable alternative to accelerator technology \cite{gravitycosmic}. Unlike in collider experiments, however, the luminosity of incoming particles is often not well known in astroparticle physics experiments. For this reason, cross section measurements cannot be inferred by the observed rate alone. At very high energies (above $\,\sim$100 TeV), however, the Earth becomes opaque to neutrinos, and the resulting angular distribution of events can be used to resolve the neutrino's interaction cross section. Using the Earth as a ``neutrino filter'' may allow kilometer-scale experiments with energy and angular resolution to effectively measure the neutrino-nucleon interaction in the range of $\sim$10 TeV to $\sim$100 PeV \cite{measure}.

Using this method can effectively probe many low-scale gravity models to a scale of $\sim$1 TeV in a kilometer scale experiment such as IceCube \cite{gravity}. The prospects for the observation of TeV string excitations are considerably enhanced if not only neutrino-quark, but also neutrino-gluon scattering amplitude, is taken into account \cite{gluons}. Microscopic black hole production can be tested effectively in both neutrino telescopes \cite{bhcosmic} and air shower experiments \cite{bhairshower}. In addition to measuring the cross section for black hole production, experiments such as IceCube are capable of measuring the ratios of muon, shower and tau-unique events and comparing them to the predictions for black hole evaporation via Hawking radiation \cite{bhcosmic}.

\section{Searches For Particle Dark Matter}

Next, I will briefly review two very different classes of dark matter candidates and two very different corresponding methods for their detection using neutrino astronomy. The first of these are TeV-scale WIMPs, such as the lightest neutralino in models of supersymmetry \cite{susydark}, or Kaluza-Klein excitations of the photon in models of universal extra dimensions \cite{kkdark}. The second class of dark matter I will discuss are superheavy particles, at or near the GUT scale.

\subsection{TeV Scale WIMPs}

If the dark matter in our galactic halo consists of TeV-scale particles, then millions of such particles travel through each square meter of our solar system each second. Although such particles are likely to have very small scattering cross sections, over long timescales, many may scatter off of bodies such as the Sun or Earth and become trapped in these deep gravitational wells. As they accumulate in the center of these bodies, their annihilation rate will be enhanced, in some cases reaching equilibrium between their capture and annihilation rates.

WIMPs could annihilate into a wide variety of channels. For example, in supersymmetry, neutralinos often annihilate to b quark or tau lepton pairs for the case of a gaugino-like neutralino or to gauge bosons for the higgsino-like case. These particles then fragment producing a spectrum of stable particles including photons, electrons, protons and neutrinos. Of these, only neutrinos can escape from the center of the Sun or Earth, potentially providing a useful signature of particle dark matter \cite{indirectneutrino}.

\begin{figure}[t]
\begin{center}
\includegraphics[width=10.0cm,angle=0]{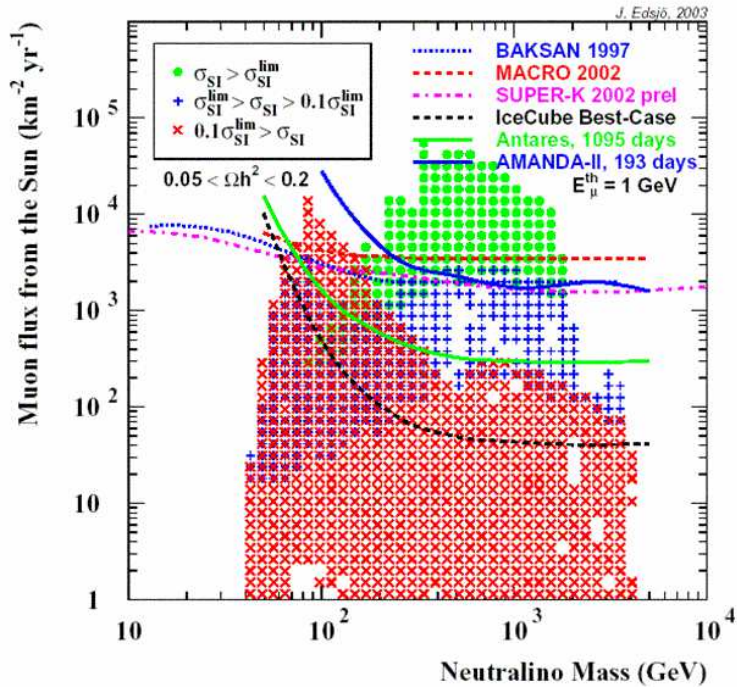}
\end{center}
\vspace*{0mm}
\caption[]{\label{susywimp}
Limits for neutralino dark matter from current neutrino telescopes, including AMANDA-II as well as the projected reach of Antares and IceCube. Notice the correlation between direct scattering sensitivity and the rates in neutrino experiments.}
\end{figure}

In figure~\ref{susywimp}, we show the sensitivity of neutrino telescopes to supersymmetric dark matter. Currently, BAKSAN, SUPER-K and AMANDA-II have similar sensitivities for WIMPs with masses above 200 GeV. The limit shown for AMANDA corresponds to only 193 days of data and should be dramatically improved upon the analysis of their full data set. Also shown in the figure are the projected sensitivities of Antares and IceCube, which each represent major improvements for WIMP searches.

There is a clear correlation between elastic scattering and neutrino telescope sensitivity to dark matter. Figure~\ref{susywimp} shows that those models probed by AMANDA, BAKSAN and SUPER-K also have spin-independent scattering cross sections above the current experimental limits. In this respect, direct dark matter searches and neutrino telescopes provide a cross check for any evidence of dark matter which may be claimed by either technique.

An alternative dark matter candidate at the TeV scale are stable Kaluza-Klein (KK) excitations which arise in models of universal extra dimensions. There are a few key phenomenological differences between this case and supersymmetric dark matter. First, such a dark matter candidate cannot be as light as sometimes can be the case in supersymmetry. In particular, electroweak precision measurements place the lightest KK particles at about 300 GeV or heavier \cite{kkbound}. Second, KK dark matter annihilates largely into charged leptons, providing a rich source of neutrinos from tau decay. Finally, unlike in supersymmetry, a direct annihilation channel to neutrinos also exists, and although only a few percent of all annihilations directly produce neutrinos, these neutrinos can be a major contributor to the experimental rate. Using estimations of the KK spectrum from calculations of the one-loop radiative corrections \cite{radiative}, annual rates of several tens of events from the Sun per square kilometer are predicted \cite{kribs}. IceCube can thus provide an effective probe of KK dark matter.

\subsection{Superheavy Dark Matter}

Although there are a number of theoretical reasons why TeV-scale dark matter is attractive, there is currently little or no experimental evidence for any such particle. Dark matter may be considerably different than described in the TeV-paradigm, perhaps consisting of superheavy particles, with masses at or near the scale of Grand Unification.

An important motivation for superheavy particles (or topological defects) comes from the observations of cosmic rays above the, so-called, Greisen-Zatsepin-Kuzmin (GZK) cutoff \cite{gzk}. This cutoff occurs due to protons scattering with CMB photons at a center-of-mass energy roughly equal to the mass of the $\Delta$-hadron (1.232 GeV). This corresponds to protons with energies of a few times $10^{19}$ eV in the lab frame. Above this energy, protons can propagate only 10 to 50 Mpc before losing much of their energy. Although many events above the GZK cutoff have been observed, no sources for ultra-high energy cosmic rays within this distance have yet been discovered.

Numerous solutions to this problem have been proposed, including new particles which constitute the super-GZK cosmic rays \cite{exo}, neutrinos with QCD scale cross sections \cite{qcdneu},
semi-local astrophysical sources \cite{loc} and top-down cosmic ray models\cite{top}. Top-down models are scenarios in which $10^{11}$ to $10^{16}$ GeV particles or topological defects decay or annihilate producing the cosmic rays observed above the GZK cutoff. These superheavy objects could also constitute the dark matter of the universe \cite{wim}.

Along with the photons and protons produced in top-down scenarios, other stable particles will be generated by this mechanism, including neutrinos. In fact, although the annihilation or decay modes of such superheavy objects are unknown, for a given mode, the fragmentation into stable particles can be calculated and the neutrino spectrum determined \cite{barbotcode}. Although IceCube is optimized for TeV-PeV energies, it is sensitive to neutrinos with energies up to the scale of the highest energy cosmic rays \cite{alvarez}. Auger and IceCube could each observe tens or even hundreds of neutrino events per year in top-down cosmic ray scenarios \cite{topneutrinos}.

\begin{figure}[t]
\begin{center}
\includegraphics[width=10.0cm,angle=0]{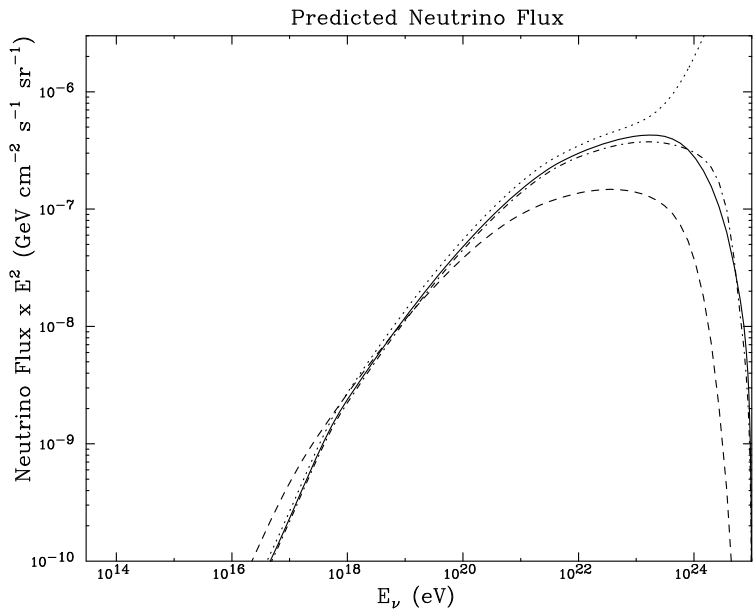}
\end{center}
\vspace*{0mm}
\caption[]{\label{topdownneutrinos}
The spectrum of very high-energy neutrinos predicted in top-down cosmic ray scenarios \cite{topneutrinos}.}
\end{figure}

In addition to neutrinos, stable supersymmetric particles may accompany the protons and photons of a top-down scenario. Although IceCube and Auger will be unlikely to be sensitive to such a signature, future satellite-based (or space station based) cosmic ray experiments, such as EUSO, may test some models \cite{toplsps}. Ultra-high energy supersymmetric particles would be a ``smoking gun'' for cosmic rays of top-down origin.

\section{Neutrino Properties}

Thus far, I have limited my discussion to using neutrinos as a tool for probing other aspects of particle physics. It is only natural that neutrino astronomy can also provide useful insights into the properties of neutrinos themselves. 

\subsection{Neutrino Decay?}

In a generic cosmic accelerator, neutrinos are produced in the decay of charged pions, producing the ratio of neutrino flavors: $\phi_{\nu_e}:\phi_{\nu_{\mu}}:\phi_{\nu_{\tau}} \simeq 1:2:0$. Over long baselines, however, oscillations change this ratio to $\phi_{\nu_e}:\phi_{\nu_{\mu}}:\phi_{\nu_{\tau}} \simeq 1:1:1$. This prediction is rather robust in the absence of exotic physics. If one or more neutrino mass eigenstates can decay over very long baselines, it can alter these flavor ratios, providing an opportunity for observation in neutrino telescopes. Current limits on neutrino decay from solar neutrino experiments restrict $\tau_{\nu}/m_{\nu} \gsim 10^{-4}\,\rm{s/eV}$. Neutrinos of 100 TeV from a 100 Mpc baseline could, in principle, provide a sensitivity $10^6$ times stronger \cite{decay}.

The details of a neutrino decay scenario can be quite varied. Similar phenomenological features appear generically, however. Table 1 shows the predicted neutrino flavor ratios for several decay scenarios. Notice that in all of the cases, the numbers of muon and tau neutrinos are equal, with only the comparative number of electron neutrinos varying. 

\begin{table}[ht]
\label{ratios}
\begin{tabular}{c c c c}
\hline
Unstable & Daughters & Branchings &  
$\phi_{\nu_e}:\phi_{\nu_\mu}:\phi_{\nu_\tau}$   \\
\hline\hline
$\nu_2$, $\nu_3$  & anything & irrelevant & $6:1:1$ \\
\hline
$\nu_3$           & sterile  & irrelevant & $2:1:1$ \\
\hline
$\nu_3$          & full energy & $B_{3 \rightarrow 2}=1$  & $1.4:1:1$ \\
                 & degraded ($\alpha=2$)               &  & $1.6:1:1$ \\
\hline
$\nu_3$          & full energy & $B_{3 \rightarrow 1}=1$   & $2.8:1:1$ \\  
                 & degraded ($\alpha=2$)               &   & $2.4:1:1$ \\
\hline
$\nu_3$          & anything & $B_{3 \rightarrow 1}=0.5$    & $2:1:1$ \\
                 &  &  $B_{3 \rightarrow 2}=0.5$           & \\  
\hline
\end{tabular}
\caption{Neutrino flavor ratios for various decay scenarios.}
\end{table}

The practical question, of course, is whether the flavor ratios of cosmic neutrinos can be accurately measured in current or planned experiments. To answer this question, I will focus on the IceCube experiment, due to its three channel measurement ability (muons, showers and tau-unique events). Using the ratios of events in these channels, one can infer the incoming flavor ratios. Furthermore, by taking advantage of the symmetry between tau and muon neutrino fluxes described above, one can use the ratio of muon tracks to shower events to infer the electron neutrino to muon neutrino ratio and use any tau-unique events which are observed as a verification of the result. For the details and prospects for neutrino flavor ratio measurements in IceCube, see Ref.~\cite{flavormeasure}.

\subsection{Pseudo-Dirac Neutrinos}

If neutrinos are pseudo-Dirac states, where each generation is composed of two maximally mixed Majorana neutrinos, separated by a very small mass difference, standard neutrino mixing phenomenology would remain unchanged from the standard scenario for $\delta m^2 \lsim 10^{-12} \, \rm{eV}^2$. Furthermore, neutrinoless double beta decay would be highly suppressed, making such small splittings very difficult to probe experimentally.

Over 100 Mpc baselines, however, even very small pseudo-Dirac mass splitting can begin to change the phenomenology of neutrino oscillations. Again, deviations from $\phi_{\nu_e}:\phi_{\nu_{\mu}}:\phi_{\nu_{\tau}} \simeq 1:1:1$ could be the signature for new physics beyond the reach of collider experiments \cite{pseudodirac}.

Unfortunately, the deviations from $\phi_{\nu_e}:\phi_{\nu_{\mu}}:\phi_{\nu_{\tau}} \simeq 1:1:1$ predicted for pseudo-Dirac neutrinos are not as dramatic as in the case of neutrino decay. It will be a challenge for next generation neutrino telescopes to reach adequate precision to test this scenario.

\subsection{Absolute Neutrino Masses}

Although neutrino mixing measurements have revealed to us that neutrinos do in fact have small masses, oscillation experiments only measure the squares of the neutrino mass splittings, leaving the absolute masses unknown. To study absolute neutrino masses, one can consider ultra-high energy cosmic neutrinos.

Extremely high-energy neutrinos propagating through the universe can interact with neutrinos of the cosmic neutrino background, in particular if their center-of-mass energy is near the Z-resonance. For an eV-scale neutrino mass, the Z-resonance corresponds to a cosmic neutrino with an energy of a few times $10^{21} \, \rm{eV}$. Such interactions would produce products of photons, protons and neutrinos following the standard spectrum of Z decays.

If cosmic sources of $10^{21}-10^{22} \rm{eV}$ neutrinos exist, it has been proposed that such neutrinos could propagate from cosmological distances, interacting and producing ultra-high energy protons (or photons) within the GZK radius, thus solving the ultra-high energy cosmic ray problem. This so-called ``Z-Burst'' mechanism \cite{zburst}, also provides a method for measuring eV-scale neutrino masses.
 
Although well out of the reach of current experiments, perhaps future ultra-high energy neutrino experiments will be successful in measuring the spectrum of cosmic neutrinos in the range above $10^{21}\rm{eV}$. Absorption lines in this spectrum could correspond to Z-resonance interactions, allowing for a new technique to measure neutrino masses: ultra-high energy neutrino spectroscopy \cite{spec}.

\section{Conclusions}

The possibilities for meaningful advances for particle physics using high-energy neutrino astronomy are numerous. In addition to observing interactions beyond the energies accessible at colliders, cosmic neutrinos can travel over hundreds or thousands of megaparsecs, thus providing new, and more stringent, tests of neutrino decay and pseudo-Dirac mass splittings. Even probing the GUT scale is a possibility for neutrino astronomy in top-down cosmic ray scenarios.

Many of the techniques described in this talk are complementary to colliders. It is clear that collider physics has substantial advantages over astroparticle physics in some respects. Accelerators offer controlled and high-luminosity environments. Collider experiments are also limited in energy and baseline, however. By taking advantage of the complementarity of collider and astroparticle physics experiments, much more progress can be made than by either alone.

\acknowledgements
I would like to thank the organizers of the 2004 Cracow Epiphany Conference for planning an informative and interesting program and for their hospitality. I would also like to thank Francis Halzen and Ignacio de la Calle Perez for helpful comments. DH is supported by the Leverhulme trust.


\newpage



\begin{thebibliography}{99}

\bibitem{amanda}
E.~Andres {\it et al.},
Nature {\bf 410}, 441 (2001).

\bibitem{antares}
F.~Blanc {\it et al.}  [ANTARES Collaboration],
{\it Presented by L. Thompson on behalf of the ANTARES Collaboration, to appear in the proceedings of 28th International Cosmic Ray Conferences
(ICRC 2003), Tsukuba, Japan, 31 Jul - 7 Aug 2003.}
U.~F.~Katz  [The ANTARES Collaboration],
arXiv:astro-ph/0310736.

\bibitem{icecube}
Ahrens,~J., {\it et al.}, 2003,
to be published in Particle Astrophysics, astro-ph/0305196; 
 IceCube Preliminary Design Document, available at:
http://icecube.wisc.edu/science/sci-tech-docs/.

\bibitem{doublebang}
J.~G.~Learned and S.~Pakvasa,
Astropart.\ Phys.\  {\bf 3}, 267 (1995)
[arXiv:hep-ph/9405296];
F.~Halzen and D.~Saltzberg,
Phys.\ Rev.\ Lett.\  {\bf 81}, 4305 (1998)
[arXiv:hep-ph/9804354].



\bibitem{rice}
I.~Kravchenko,
arXiv:astro-ph/0306408;
I.~Kravchenko {\it et al.},
Astropart.\ Phys.\  {\bf 20}, 195 (2003)
[arXiv:astro-ph/0206371].


\bibitem{anita}
ANITA Proposal, P.~Gorham, Principle Investigator; ANITA Home Page: http://www.ps.uci.edu/~anita/index.html.



\bibitem{radio}
I.~Kravchenko,
arXiv:astro-ph/0306408.

\bibitem{acoustic}
L.~G.~Dedenko, I.~M.~Zheleznykh, S.~K.~Karaevsky, A.~A.~Mironovich,
V.~D.~Svet and A.~V.~Furduev,
Bull.\ Russ.\ Acad.\ Sci.\ Phys.\  {\bf 61} (1997) 469
[Izv.\ Ross.\ Akad.\ Nauk.\  {\bf 61} (1997) 593].


\bibitem{auger}
J.~Blumer  [AUGER Collaboration],
{\it Prepared for 28th International Cosmic Ray Conferences (ICRC 2003), Tsukuba, Japan, 31 Jul - 7 Aug 2003};
S.~Coutu, X.~Bertou and P.~Billoir  [AUGER Collaboration],
{\it Prepared for 23rd Johns Hopkins Workshop on Current Problems in Particle Theory: Neutrinos in the New Millennium, Baltimore, Maryland, 10-12
Jun 1999};
K.~S.~Capelle, J.~W.~Cronin, G.~Parente and E.~Zas,
Astropart.\ Phys.\  {\bf 8}, 321 (1998)
[arXiv:astro-ph/9801313].



\bibitem{euso}
L.~Scarsi,
Nuovo Cim.\  {\bf 24C}, 471 (2001);
D.~B.~Cline and F.~W.~Stecker,
arXiv:astro-ph/0003459.




\bibitem{salsa}
P.~Gorham, D.~Saltzberg, A.~Odian, D.~Williams, D.~Besson, G.~Frichter and S.~Tantawi,
Nucl.\ Instrum.\ Meth.\ A {\bf 490}, 476 (2002)
[arXiv:hep-ex/0108027].

\bibitem{hypercube}
F.~Halzen and D.~Hooper,
JCAP/008A/1003, arXiv:astro-ph/0310152.

\bibitem{review}
F.~Halzen and D.~Hooper,
Rept.\ Prog.\ Phys.\  {\bf 65}, 1025 (2002)
[arXiv:astro-ph/0204527].


\bibitem{add}
N.~Arkani-Hamed, S.~Dimopoulos and G.~R.~Dvali,
Phys.\ Lett.\ B {\bf 429}, 263 (1998)
[arXiv:hep-ph/9803315];
I.~Antoniadis, N.~Arkani-Hamed, S.~Dimopoulos and G.~R.~Dvali,
Phys.\ Lett.\ B {\bf 436}, 257 (1998)
[arXiv:hep-ph/9804398].

\bibitem{rs}
L.~Randall and R.~Sundrum,
Phys.\ Rev.\ Lett.\  {\bf 83}, 3370 (1999)
[arXiv:hep-ph/9905221].

\bibitem{veneziano}
K.~Benakli,
Phys.\ Rev.\ D {\bf 60}, 104002 (1999)
[arXiv:hep-ph/9809582];
E.~Accomando, I.~Antoniadis and K.~Benakli,
Nucl.\ Phys.\ B {\bf 579}, 3 (2000)
[arXiv:hep-ph/9912287];
S.~Cullen, M.~Perelstein and M.~E.~Peskin,
Phys.\ Rev.\ D {\bf 62}, 055012 (2000)
[arXiv:hep-ph/0001166].




\bibitem{bh}
D.~Amati, M.~Ciafaloni and G.~Veneziano,
Phys.\ Lett.\ B {\bf 197}, 81 (1987);
D.~Amati, M.~Ciafaloni and G.~Veneziano,
Phys.\ Lett.\ B {\bf 289}, 87 (1992);
G.~'t Hooft,
Phys.\ Lett.\ B {\bf 198}, 61 (1987);
P.~C.~Argyres, S.~Dimopoulos and J.~March-Russell,
Phys.\ Lett.\ B {\bf 441}, 96 (1998)
[arXiv:hep-th/9808138];
T.~Banks and W.~Fischler,
arXiv:hep-th/9906038;
R.~Emparan, G.~T.~Horowitz and R.~C.~Myers,
Phys.\ Rev.\ Lett.\  {\bf 85}, 499 (2000)
[arXiv:hep-th/0003118];
S.~B.~Giddings and S.~Thomas,
Phys.\ Rev.\ D {\bf 65}, 056010 (2002)
[arXiv:hep-ph/0106219].


\bibitem{instanton}
G.~'t Hooft,
Phys.\ Rev.\ Lett.\  {\bf 37}, 8 (1976);
G.~'t Hooft,
Phys.\ Rev.\ D {\bf 14}, 3432 (1976)
[Erratum-ibid.\ D {\bf 18}, 2199 (1978)];
A.~Ringwald,
Nucl.\ Phys.\ B {\bf 330}, 1 (1990);
V.~V.~Khoze and A.~Ringwald,
Phys.\ Lett.\ B {\bf 259}, 106 (1991);
Z.~Fodor, S.~D.~Katz, A.~Ringwald and H.~Tu,
Phys.\ Lett.\ B {\bf 561}, 191 (2003)
[arXiv:hep-ph/0303080];
A.~Ringwald,
Phys.\ Lett.\ B {\bf 555}, 227 (2003)
[arXiv:hep-ph/0212099].




\bibitem{gravitycosmic}
A.~Jain, P.~Jain, D.~W.~McKay and J.~P.~Ralston,
arXiv:hep-ph/0011310;
J.~P.~Ralston, P.~Jain, D.~W.~McKay and S.~Panda,
AIP Conf.\ Proc.\  {\bf 549}, 733 (2002)
[arXiv:hep-ph/0008153].
P.~Jain, D.~W.~McKay, S.~Panda and J.~P.~Ralston,
Phys.\ Lett.\ B {\bf 484}, 267 (2000)
[arXiv:hep-ph/0001031].

\bibitem{measure}
D.~Hooper,
Phys.\ Rev.\ D {\bf 65}, 097303 (2002)
[arXiv:hep-ph/0203239].

\bibitem{gravity}
J.~Alvarez-Muniz, F.~Halzen, T.~Han and D.~Hooper,
Phys.\ Rev.\ Lett.\  {\bf 88}, 021301 (2002)
[arXiv:hep-ph/0107057].



\bibitem{gluons}
J.~J.~Friess, T.~Han and D.~Hooper,
Phys.\ Lett.\ B {\bf 547}, 31 (2002)
[arXiv:hep-ph/0204112].



\bibitem{bhcosmic}
M.~Kowalski, A.~Ringwald and H.~Tu,
Phys.\ Lett.\ B {\bf 529}, 1 (2002)
[arXiv:hep-ph/0201139];
J.~Alvarez-Muniz, J.~L.~Feng, F.~Halzen, T.~Han and D.~Hooper,
Phys.\ Rev.\ D {\bf 65}, 124015 (2002)
[arXiv:hep-ph/0202081].




\bibitem{bhairshower}
J.~L.~Feng and A.~D.~Shapere,
Phys.\ Rev.\ Lett.\  {\bf 88}, 021303 (2002)
[arXiv:hep-ph/0109106];
L.~A.~Anchordoqui, J.~L.~Feng, H.~Goldberg and A.~D.~Shapere,
Phys.\ Rev.\ D {\bf 68}, 104025 (2003)
[arXiv:hep-ph/0307228];
L.~A.~Anchordoqui, J.~L.~Feng, H.~Goldberg and A.~D.~Shapere,
Phys.\ Rev.\ D {\bf 65}, 124027 (2002)
[arXiv:hep-ph/0112247];
L.~Anchordoqui and H.~Goldberg,
Phys.\ Rev.\ D {\bf 65}, 047502 (2002)
[arXiv:hep-ph/0109242].


\bibitem{instantoncosmic}
T.~Han and D.~Hooper,
arXiv:hep-ph/0307120.




\bibitem{susydark}
H.~Goldberg, Phys.\ Rev.\ Lett.\ {\bf 50}, 1419 (1983);
%
J.~Ellis, J.~S.~Hagelin, D.~V.~Nanopoulos, K.~Olive and M.~Srednicki,
Nucl.\ Phys.\ B {\bf 238}, 453 (1984).




\bibitem{kkdark}
E.~W.~Kolb and R.~Slansky,
Phys.\ Lett.\ B {\bf 135}, 378 (1984);
K.~R.~Dienes, E.~Dudas and T.~Gherghetta,
Nucl.\ Phys.\ B {\bf 537}, 47 (1999)
[arXiv:hep-ph/9806292];
G.~Servant and T.~M.~Tait,
arXiv:hep-ph/0206071.
H.~C.~Cheng, J.~L.~Feng and K.~T.~Matchev,
arXiv:hep-ph/0207125.




\bibitem{indirectneutrino}
J.~Silk, K.~Olive and M.~Srednicki,
Phys.\ Rev.\ Lett.\ {\bf 55}, 257 (1985); 
J.~S.~Hagelin, K.~W.~Ng, K.~A.~Olive, 
Phys.\ Lett.\ B 180:375 (1986);
L.~M.~Krauss, M.~Srednicki and F.~Wilczek,
Phys.\ Rev.\ D {\bf 33}, 2079 (1986);
T.~K.~Gaisser, G.~Steigman and S.~Tilav,
Phys.\ Rev.\ D {\bf 34}, 2206 (1986);
G.~Jungman and M.~Kamionkowski,
Phys.\ Rev.\ D {\bf 51}, 328 (1995)
[arXiv:hep-ph/9407351];
V.~Berezinsky, A.~Bottino, J.~Ellis, N.~Fornengo, G.~Mignola and S.~Scopel,
Astropart.\ Phys.\ {\bf 5}, 333 (1996)
[hep-ph/9603342];
L.~Bergstrom, J.~Edsjo and P.~Gondolo,
Phys.\ Rev.\ D {\bf 55}, 1765 (1997);
Phys.\ Rev.\ D {\bf 58}, 103519 (1998);
L.~Bergstrom, J.~Edsjo and M.~Kamionkowski,
Astropart.\ Phys.\ {\bf 7}, 147 (1997);
A.~Corsetti and P.~Nath,
Int.\ J.\ Mod.\ Phys.\ A {\bf 15}, 905 (2000);
A.~E.~Faraggi, K.~A.~Olive and M.~Pospelov,
Astropart.\ Phys.\ {\bf 13}, 31 (2000);
J.~L.~Feng, K.~T.~Matchev and F.~Wilczek,
Phys.\ Rev.\ D {\bf 63}, 045024 (2001);
V.~D.~Barger, F.~Halzen, D.~Hooper and C.~Kao,
Phys.\ Rev.\ D {\bf 65}, 075022 (2002);
P.~Crotty,
arXiv:hep-ph/0205116.
D.~Hooper and L.~T.~Wang,
arXiv:hep-ph/0309036;




\bibitem{kkbound}
T.~Appelquist, H.~C.~Cheng and B.~A.~Dobrescu,
Phys.\ Rev.\ D {\bf 64}, 035002 (2001)
[arXiv:hep-ph/0012100].

\bibitem{radiative}
H.~C.~Cheng, K.~T.~Matchev and M.~Schmaltz,
arXiv:hep-ph/0204342;
arXiv:hep-ph/0205314.


\bibitem{kribs}
D.~Hooper and G.~D.~Kribs,
Phys.\ Rev.\ D {\bf 67}, 055003 (2003)
[arXiv:hep-ph/0208261].





\bibitem{gzk}
K.~Greisen,
Phys.\ Rev.\ Lett.\  {\bf 16}, 748 (1966); G.~T.~Zatsepin and
V.~A.~Kuzmin,
JETP Lett.\  {\bf 4}, 78 (1966)
[Pisma Zh.\ Eksp.\ Teor.\ Fiz.\  {\bf 4}, 114 (1966)];
For the up-to-date status of ultra-high energy cosmic rays, see the
AGASA Homepage, www.icrr.u-tokyo.ac.jp/as/as.html and the
HIRES Homepage, www2.keck.hawaii.edu:3636/realpublic/inst/hires/hires.html.


\bibitem{exo}
A.~Perez-Lorenzana,
{\it  In Oaxaca de Juarez 1998, Particles and fields 409-412};
L.~Masperi,
{\it  In Trieste 1998, Non-accelerator particle astrophysics 218-224};
I.~F.~Albuquerque, G.~R.~Farrar and E.~W.~Kolb,
Phys.\ Rev.\ D {\bf 59}, 015021 (1999),
hep-ph/9805288.


\bibitem{qcdneu}
A.~Jain, P.~Jain, D.~W.~McKay and J.~P.~Ralston,
hep-ph/0011310; P.~Jain, D.~W.~McKay, S.~Panda and J.~P.~Ralston,
Phys.\ Lett.\ B {\bf 484}, 267 (2000), hep-ph/0001031; G.~Domokos,
S.~Kovesi-Domokos and P.~T.~Mikulski,
hep-ph/0006328; M.~Kachelriess and M.~Pl\"umacher,
Phys.\ Rev.\ D {\bf 62}, 103006 (2000), astro-ph/0005309.



\bibitem{loc}
C.~Isola, M.~Lemoine and G.~Sigl,
astro-ph/0104289; P.~Blasi and A.~V.~Olinto,
Phys.\ Rev.\ D {\bf 59}, 023001 (1999), astro-ph/9806264.



\bibitem{top}
R.~J.~Protheroe and T.~Stanev,
Phys.\ Rev.\ Lett.\  {\bf 77}, 3708 (1996), astro-ph/9605036; 
G.~Sigl, S.~Lee, P.~Bhattacharjee and S.~Yoshida,
Phys.\ Rev.\ D {\bf 59}, 043504 (1999), hep-ph/9809242;
P.~Bhattacharjee, C.~T.~Hill and D.~N.~Schramm,
Phys.\ Rev.\ Lett.\  {\bf 69}, 567 (1992);
S.~Sarkar and R.~Toldra,
Nucl.\ Phys.\ B {\bf 621}, 495 (2002)
[arXiv:hep-ph/0108098];
S.~Sarkar,
Astropart.\ Phys.\  {\bf 9}, 297 (1998), hep-ph/9804285;
V.~Berezinsky, M.~Kachelriess and A.~Vilenkin,
Phys.\ Rev.\ Lett.\  {\bf 79}, 4302 (1997), astro-ph/9708217;
O.E. Kalashev, V.A. Kuzmin, D.V. Semikoz and G. Sigl,
Phys.\ Rev.\ D {\bf 66}, 063004 (2002),
[arXiv:hep-ph/0205050]. 


\bibitem{wim}
D.~J.~Chung, E.~W.~Kolb and A.~Riotto,
Phys.\ Rev.\ D {\bf 59}, 023501 (1999), hep-ph/9802238;
P.~Blasi, R.~Dick and E.~W.~Kolb,
astro-ph/0105232; D.~J.~Chung, P.~Crotty, E.~W.~Kolb and A.~Riotto,
Phys.\ Rev.\ D {\bf 64}, 043503 (2001), hep-ph/0104100; K.~Benakli, J.~R.~Ellis
and D.~V.~Nanopoulos,
Phys.\ Rev.\ D {\bf 59}, 047301 (1999), hep-ph/9803333; V.~Berezinsky,
M.~Kachelriess, and A.~Vilenkin,
Phys.\ Rev.\ Lett.\  {\bf 79}, 4302 (1997); C.~Coriano, A.~E.~Faraggi
and M.~Pl\"umacher,
Nucl.\ Phys.\ B {\bf 614}, 233 (2001), hep-ph/0107053; S.~Chang,
C.~Coriano and A.~E.~Faraggi,
Nucl.\ Phys.\ B {\bf 477}, 65 (1996), hep-ph/9605325; H.~Ziaeepour,
Astropart.\ Phys.\  {\bf 16}, 101 (2001), astro-ph/0001137; R.~Allahverdi and M.~Drees, hep-ph/0203118..






\bibitem{barbotcode}
C.~Barbot and M.~Drees,
Astropart.\ Phys.\  {\bf 20}, 5 (2003)
[arXiv:hep-ph/0211406];
C. Barbot and M. Drees, Phys. Lett. {\bf B533}, 107 (2002),
hep--ph 0202072.

\bibitem{alvarez}
J.~Alvarez-Muniz and F.~Halzen,
Phys.\ Rev.\ D {\bf 63}, 037302 (2001)
[arXiv:astro-ph/0007329];
J.~Alvarez-Muniz and F.~Halzen,
AIP Conf.\ Proc.\  {\bf 579}, 305 (2001)
[arXiv:astro-ph/0102106].

\bibitem{topneutrinos}
C.~Barbot, M.~Drees, F.~Halzen and D.~Hooper,
Phys.\ Lett.\ B {\bf 555}, 22 (2003)
[arXiv:hep-ph/0205230].


\bibitem{toplsps}
C.~Barbot, M.~Drees, F.~Halzen and D.~Hooper,
Phys.\ Lett.\ B {\bf 563}, 132 (2003)
[arXiv:hep-ph/0207133].




\bibitem{decay}
J.~F.~Beacom, N.~F.~Bell, D.~Hooper, S.~Pakvasa and T.~J.~Weiler,
Phys.\ Rev.\ Lett.\  {\bf 90}, 181301 (2003)
[arXiv:hep-ph/0211305];
J.~F.~Beacom, N.~F.~Bell, D.~Hooper, S.~Pakvasa and T.~J.~Weiler,
arXiv:hep-ph/0309267.


\bibitem{flavormeasure}
J.~F.~Beacom, N.~F.~Bell, D.~Hooper, S.~Pakvasa and T.~J.~Weiler,
arXiv:hep-ph/0307025.


\bibitem{pseudodirac}
J.~F.~Beacom, N.~F.~Bell, D.~Hooper, J.~G.~Learned, S.~Pakvasa and T.~J.~Weiler,
arXiv:hep-ph/0307151.




\bibitem{zburst}
T.~Weiler,
Phys.\ Rev.\ Lett.\  {\bf 49}, 234 (1982);
T.~J.~Weiler,
Astropart.\ Phys.\  {\bf 11}, 303 (1999), hep-ph/9710431;
D.~Fargion, B.~Mele and A.~Salis,
Astrophys.\ J.\  {\bf 517}, 725 (1999), astro-ph/9710029;
Z.~Fodor, S.~D.~Katz and A.~Ringwald,
Phys.\ Rev.\ Lett.\  {\bf 88}, 171101 (2002)
[arXiv:hep-ph/0105064];
Z.~Fodor, S.~D.~Katz and A.~Ringwald,
JHEP {\bf 0206}, 046 (2002)
[arXiv:hep-ph/0203198].


\bibitem{spec}
A.~Ringwald, T.~Weiler, L.~Song, Eberle,
In preparation.


\end{thebibliography}
\end{document}